\documentclass[%
 aip,
 jmp,%
 amsmath,amssymb,
 reprint,%
]{revtex4-1}
\usepackage{graphicx}
\usepackage{dcolumn}
\usepackage{bm}

\newcommand{\be}{\begin{equation}}
\newcommand{\ee}{\end{equation}}
\newcommand{\bea}{\begin{eqnarray}}
\newcommand{\eea}{\end{eqnarray}}

\begin{document}

\preprint{AIP/123-QED}

\title{Light-induced long-ranged disorder effect in ultra-dilute two-dimensional holes in GaAs heterojunction-insulated-gate field-effect-transistors}
\thanks{email:jianhuang@wayne.edu}

\author{Jian Huang}
\affiliation{%
Department of Physics and Astronomy, Wayne State University, Detroit, MI 48201, USA\\}%
\affiliation{%
Department of Electrical Engineering, Princeton University, Princeton, NJ 08544}%
\author{L. N. Pfeiffer}%
\author{K. W. West}%
\affiliation{%
Department of Electrical Engineering, Princeton University, Princeton, NJ 08544}%

\date{\today}

\begin{abstract}
Comparing the results of transport measurements of strongly correlated two-dimensional holes in a GaAs heterojunction-insulated-gate field-effect-transistor obtained before and after a brief photo-illumination, the light-induced disorder is found to cause qualitative changes suggesting altered carrier states. For charge concentrations ranging from $3\times10^{10}$ $cm^{-2}$ down to $7\times10^{8}$ cm$^{-2}$, the post-illumination hole mobility exhibits a severe suppression for charge densities below $2\times10^{10}$ cm$^{-2}$, while almost no change for densities above. The long-ranged nature of the disorder is identified. The temperature dependence of the conductivity is also drastically modified by the disorder reconfiguration from being nonactivated to activated.
\end{abstract}

\pacs{Valid PACS appear here}
\maketitle

Low temperature charge transport study of strongly correlated two-dimensional (2D) electron
systems is an important means of probing electron-disorder scattering and electron-electron interaction, both of which are fundamentally important. In 2D semiconductor systems, the level of quenched disorder is inherent and may vary drastically from sample to sample depending on the material, growth conditions, and the processing methods. Generally, disorder dominated electron systems exhibit insulating behaviors that are mostly understood through the Anderson Localization~\cite{Anderson'58}. The finite temperature conductivity of such systems is phonon-activated: \be \label{expo} {\sigma(T)} \sim e^{(T^*/T) ^\mu} \,, \quad \mu = 1/3, 1/2, 1 \ee $\mu=1$ for Arrhenius conductivity and $\nu=1/3$ and $\nu=1/2$ are for Mott~\cite{Mott-VRH} and Efros-Shklovskii variable range hopping (VRH)~\cite{ES}. On the other hand, strong electron-electron interaction is also expected to qualitatively alter the states of the electrons, or even cause phase transitions, i.e. Wigner crystallization, when the interaction becomes dominant~\cite{wc,wc1,AKS-review}. Experimental study of interaction-driven phenomena demands very high quality systems with minimal disorder, which has been a long-standing challenge. As cleaner 2D systems (i.e. GaAs and Si-MOSFET semiconductor devices) become available over time, lower charge densities are more accessible. The effective interaction becomes enhanced because the interaction parameter $r_s$
, the ratio of the Coulomb potential $E_{ee}$ and the Fermi energy $E_F$, increases with decreasing density. As a result, rising interaction effect competes with disorder, leading to unexpected phenomena such as the 2D metal-to-insulator transition (MIT)~\cite{mit}. Understanding the disorder-interaction interplay in a strongly interacting regime is a fundamental subject that has recently generated renewed interests. Two of the primary components for experiments are the capabilities of varying interaction and varying disorder independently.

\begin{figure}[t]
\vspace{0pt}
\includegraphics[totalheight=1.3in,trim=0.0in 0.0in 0.0in 0in]{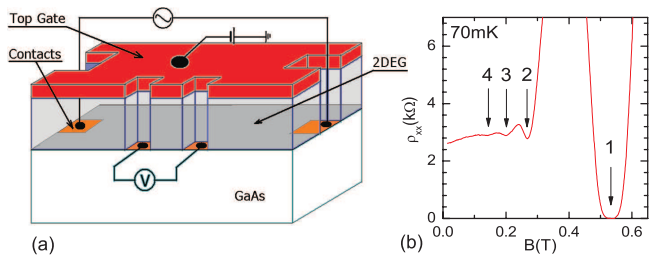}
\vspace{-5pt}
\caption{\label{fig:rx} (Color online) (a) HIGFET structure and measurement schematics. (b) $R_{xx}$ vs.  B-field at $T=70mK$.}
\vspace{-15pt}
\end{figure}

Early studies mainly focused on the changes of transport characteristics caused by varying charge densities (or interaction), while disorder is fixed. In order to investigate whether strong interaction causes qualitative changes to the charge states, the cleanest systems available were used. Even so, most results on the insulating side of MIT show activated conduction, signifying disorder domination. Therefore, higher purity systems are desired. Lately, an unique type of higher purity undoped GaAs/AlGaAs heterojunction-insulated-gate field-effect-transistor (HIGFET) has been adopted. The charges are capacitively induced at the GaAs/AlGaAs 2D interface through a metal gate separated by an AlGaAs barrier without any intentional doping (Fig.~\ref{fig:rx}(a)). Recently, ultra-dilute carrier concentrations down to $6\times10^{8}$ cm$^{-2}$ have been realized in $p$-channel HIGFETs~\cite{noh,jian-1}, corresponding to a $r_s$ beyond 40. The carrier mobility reaches $1\times10^6$ cm$^{-2}/V\cdot s$ for $p=1\times10^{10}$ cm$^{-2}$. In this Wigner Crystal regime, the transport differs from the usual activated hopping conduction~\cite{jian-1}.

An equally important yet often neglected question is how transport characteristics is affected by changing disorder, especially in a strongly correlated regime. Typically, changing disorder is realized through preparing multiple devices each having a different disorder level~\cite{wanli,long-short}. However, varying disorder within the same system is challenging. In this letter, we demonstrate a disorder reconfiguration within the same high-purity $p$-channel GaAs HIGFET system via a very brief LED (light emitting diode) illumination. Our method of varying interaction is through changing charge concentration in zero magnetic (B) field, which has the advantage of preserving the natural wavefunction form. The undoped (100) HIGFET~\cite{noh,jian-1} provides a dilute range of 2D hole charge densities from $7\times10^{10}$ cm$^{-2}$ down to $7\times10^{8}$ cm$^{-2}$. Transport measurements are made in the dark both before and after the LED illumination and drastic changes are observed. The density dependence of the hole mobility exhibits increasingly larger difference at charge densities lower than $2\times10^{10}$ cm$^{-2}$, but little changes for densities above. The long-ranged nature of the light-induced disorder is identified by considering the weakening of the charge screening at dilute charge concentrations. Meanwhile, the temperature dependence of the conductivity ($\sigma(T)$) takes on an Arrhenius activated conduction, in contrast to the power-law-like $\sigma(T)$ before the illumination.

\begin{figure}[b]
\vspace{-35pt}
\includegraphics[totalheight=2.7in,trim=0.2in 0.10in 0.20in 0in]{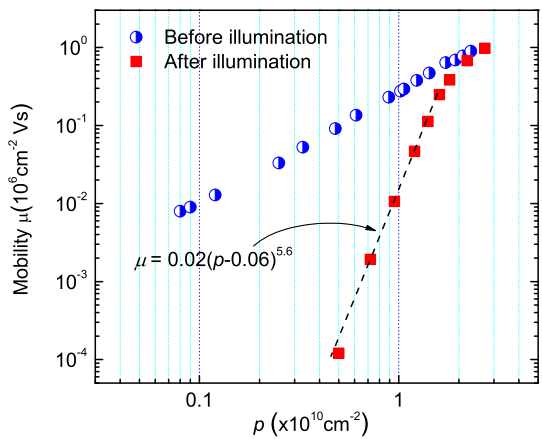}
\vspace{-15pt}
\caption{\label{fig:mb} (Color online) Mobility vs. density for before and after the illumination.}
\vspace{-10pt}
\end{figure}

The $p$-HIGFET sample preparation and measurement details are provided in Ref.~\cite{jian-4}. The measurement was made in a dilution refrigerator with the sample placed in the 3He/4He mixing chamber. The density of the 2D holes was determined by measuring the quantum oscillations of the magnetoresistance and the Hall resistance in a B-field, instead of using a fixed capacitance. Fig~\ref{fig:rx}(b) shows the $R_{xx}(B)$ from which the corresponding charge density, $1.27\times10^{10}$ $cm^{-2}$ for this case, is determined by identifying the filling factors. The measurement of the $T$-dependence of the conductivity was first made in the dark in a dilution refrigerator. Then, while the sample is held at a base temperature of 30mK, a very brief photo illumination via a red LED driven by current of $0.1\mu A$, is turned on for 1 second while the top gate of the HIGFET sample was biased at -1.2V. The generated photon energy is around 1.9-2\,eV. The threshold voltage of the FET is shifted by the illumination from -1.23\,V to -1.56\,V. After turning the device off and grounding all sample leads, there was a 24-hour waiting time before resuming the measurements in the dark and all the results collected thereafter are consistent. The pre-illumination characteristics is recovered after a cooling recycle. For the measurement in the high impedance range, dc techniques with low-level drives ({\it p}A and 0.05$\mu$V) were adopted. Results obtained from both current drive setup and voltage drive setup are in agreement.


Fig.\ref{fig:mb} shows the density-mobility ($\mu$) relationship with a post-illumination mobility (in squares) in comparison with that (in circles) measured before illumination. It should be noted that the term mobility is used in the insulating regime in a colloquial manner and should not be considered in the same sense as the metallic situations. The change of the $\mu(p)$ due to the illumination is remarkable and is significantly different than shining light in doped higher density devices which results in increased charge density and mobility. The disorder reconfiguration caused by illumination suppresses the mobility severely at low densities (below $1\times10^{10}$ $cm^{-2}$), while exerting no influence at higher densities. This is in contrast to the usual vertical shifts of the $\mu(p)$ found previously through comparing samples with different disorder concentrations. Here, the density dependence of both curves at densities below $2\times10^{10}$ $cm^{-2}$ follow approximately a power law with exponents of $\sim$1.5 before illumination and $\sim$6 after. This enormous mobility suppression only at dilute charge concentrations suggests that the light-introduced disorder is long-ranged if the disorder screening by 2D holes is taken into consideration.

At higher charge densities, long-ranged disorder potential drops faster than $1/r$ due to the screening effect of 2D charges which suppress the scattering down to short-ranged within the screening length. The screening length depends on both $q_{TF}$ (if considering small wavevectors in slowly varying disorder potential in a Thomas-Fermi case), $E_F$, as well as the strong interaction which tends to enhance the screening~\cite{Tanaskovic}. Overall, screening becomes weakened as charge density decreases which lowers $E_F$. Theoretical calculations requires many-body techniques and the density and temperature dependence of the screening length are currently unavailable. Nevertheless, at sufficient low charge densities, disorders become unscreened as the screening length is exceeded by the average charge spacing (2$a$, $a$ being the Wigner Seitz radius), resulting in maximum scattering due to the unscreened disorder. The average charge spacing varies approximately from 100nm to 400nm for the measured charge densities from 7 to $0.7\times10^{9}$ $cm^{-2}$. Such drastic change in screening should not be present if the disorder were short-ranged in nature. Thus, the change in $\mu(p)$ before and after LED illumination helps to identify the long-ranged nature of the disorder. The same mobility found for $p\geq2\times10^{10}$ $cm^{-2}$ before and after illumination indicates that the background disorder before illumination is low and short-ranged, consistent with the short-ranged nature of the impurities lying in the hetero-interface.
\begin{figure}[t]
\center{
\vspace{-30pt}
\includegraphics[width=3.6in,trim=0.3in 0.1in 0.0in 0in]{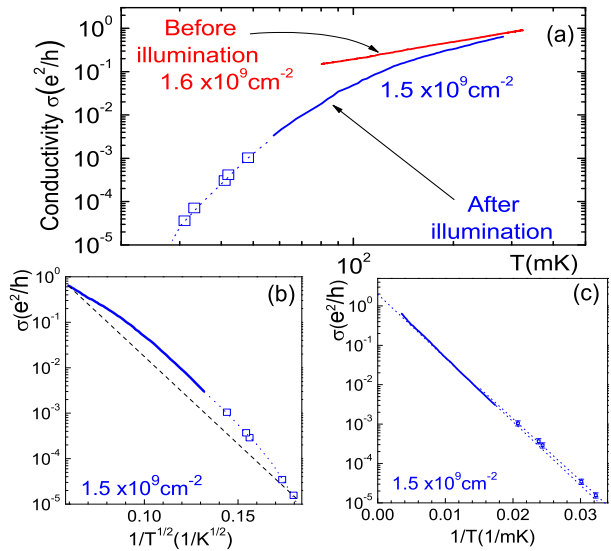}
\vspace{-30pt}
\caption{\label{fig:ct} (color online) (a) $\ln\sigma$ vs. $T$ for before- and post-illumination. The post-illumination $\sigma(T)$ compared to (b) $1/T^{1/2}$ and (c) $1/T$ for $p=1.5\times10^{9}$ cm$^{-2}$.}
}
\vspace{-15pt}
\end{figure}

The disorder reconfiguration also brings qualitative changes to the $T$-dependence of the conductivity shown in Fig.~\ref{fig:ct} where $\sigma(T)$ for before and after illumination are compared. The critical density $p_c$ of the apparent metal-to-insulator transition is roughly $\sim1\times10^{10}$ $cm^{-2}$ after illumination, significantly higher than the $p_c$ of $\sim4\times10^{9}$ cm$^{-2}$ obtained before illumination. It confirms that the long-ranged disorder is more effective in suppressing metallic behavior~\cite{long-short}.
For $p=1.6\times10^{9}$ $cm^{-2}$ in the insulating side of the MIT shown in Fig.~\ref{fig:ct}(a), $\sigma(T)$ exhibits an approximate power-law behavior before illumination~\cite{jian-1}. Notice the plot is in $lg-lg$ scales. However, for the post-illumination $\sigma(T)$ for a selected density $p=1.5\times10^{9}$\,cm$^{-2}$ (in blue), $\sigma(T)$ exhibits a much suppressed conductivity by roughly three orders of magnitudes at $T=40$\,mK. Note that the solid curve obtained from the ac measurement is connected by an artificial dotted line with the scattered points obtained from dc measurements for the same density. In addition, $\sigma(T)$ shows stronger decrease with cooling than the usual variable-range-hopping (VRH) (Fig.~\ref{fig:ct} (b)) observed in similar doped systems~\cite{mit,mitGaAs}. In Fig.~\ref{fig:ct} (c), the Arrhenius character of the $\sigma(T)$ is confirmed in the $\lg T$ vs $1/T$ plot for $T$ ranging from $\sim$500\,mK down to 30\,mK (or (1/T) from 0.002 to 0.033\,mK$^{-1}$). The linear fit in the $lg-lg$ scales intersects with y-axis ($T^*/T\ll 1$) at $\sim\sigma\sim2e^2/h$, in agreement with a previous observation obtained in doped GaAs heterostructure 2D systems~\cite{mitGaAs}.

Complimentary to a recent study intended to distinguish the contributions of short-ranged disorder from the long-ranged disorder~\cite{long-short} with two individual devices in the metallic side of MIT, our results deal with the insulating regime in the presence of strong interaction. Introducing disorder with light allows us to identify the qualitative change in transport characteristics caused only by the increase of long-ranged disorder within the same interaction range. It is worth noting that the 2D hole mobility before and after illumination reaches $\sim1\times10^6$ cm$^{-2}/V\cdot s$ for $p=2\times10^{10}$ cm$^{-2}$, signifying lower disorder levels than those in Ref.~\cite{long-short}. Prior to the LED illumination, the disorder is predominantly short-ranged, making the system an ideal candidate for an Anderson insulator. However, the non-activated power-law-like $\sigma(T)$ emerges instead of the activated $\sigma(T)\sim e^{{(T^*/T)}^{\mu}}$, contradicting a disorder-dominated scenario. After the illumination, the long-ranged disorder makes the system a better candidate for the percolation model~\cite{perco,perco-1} than for an Anderson insulator. Nevertheless, the Arrhenius conduction is somewhat unexpected due to the presence of a strong interaction~\cite{basko}. The characteristic energy $T^*$ obtained through fitting for $p=1.5\times10^{9}$ cm$^{-2}$ is roughly 395\,mK, significantly less than those found previously~\cite{mit,mitGaAs}.

Considering the large $r_s$ values, this Arrhenius behavior may also be due to a hopping conduction through imperfections in a Wigner crystal or a Wigner glass pinned by the disorder. The activation energy is then associated with the Coulomb barrier ($e^2/\epsilon a$) due to neighboring charges, as well as the smaller wavefunction overlap between localized states that decreases exponentially with increasing charge spacings.

To summarize, in a high-purity system of ultra-dilute GaAs 2D holes, we have reconfigured the disordered environment within the same sample by an LED illumination. This has resulted a qualitative change in the transport characteristics of conduction from a power-law like $T$-dependence, possibly interaction-driven, to an activated $T$-dependence controlled by disorder. The corresponding change in the density dependence of the charge mobility helps to identify the long-ranged nature of the light-induced disorder. For the Arrhenius conduction, the characteristic energy is very small due to the low disorder level present in the system. Further systematic study is need to determine how the transport varies as a function of the illumination time and the photon energy.

\bibliography{aipsamp}
\end{document}